Many-Body Quantum Calculations for Graphyne for Electronic Devices

Xianwei Sha*

General Dynamics Information Technology Corporation, Falls Church, VA 22042 and Information Technology Division, Center for Computational Science, Naval Research Laboratory, Washington, DC; *Onsite Contractor

and

Clifford M. Krowne
Electromagnetics Technology Branch, Electronics Science & Technology Division, Naval Research Laboratory, Washington, DC 20375

## Abstract

Moving beyond traditional 2D materials is now desirable to have switching capabilities (e.g., transistors). Here we propose using graphyne because, as we will show in this letter, obtaining regions of the electronic bandstructure which act as valence and conduction bands, with an apparent bandgap, is found. Here for particular allotropes of graphyne, density of states (DOS) and electronic bandstructure diagrams with $\varepsilon(\mathbf{k})$ vs $\mathbf{k}$ are found for the graphyne allotropes graphyne-1 and graphyne-2 having, respectively, one and two triple C-C carbon-carbon bonds between adjoining benzene rings. The ab initio many-body quantum calculations were performed using both local density approximation (LDA) and generalized gradient approximation (GGA) for density functional theory (DFT).

## Introduction

Carbon in a higher local energy minima state from graphene, creates a new type of carbon-based 2D material, graphyne. It shows promise as a high mobility transport material for electronics too, similar to graphene, but unlike pure graphene which has Dirac states at symmetry points in k-space and no bandgap, graphyne promises finite bandgaps. However, the intrinsic simplicity of dealing with a monoatomic material is tremendously attractive still, although the structures of graphyne are considerably more complicated than its sister material graphene. One is reminded of the initial fascination with silicon and germanium in the 1950s and 1960s. Silicon grew into the workhorse of electronics for transistors for digital computers and lower frequency RF electronics. In particular, one can think of the use of silicon for affordable cell phones and digital cameras.

However, for high frequency electronic devices, it is well known that antimonide-based compound semiconductors can have very high mobility and offer important advantages for low-power, low-noise RF electronics [2]. Unfortunately, antimonide-based materials suffer several important disadvantages including In and Sb do not have commercially-viable domestic sources [3]; InSb-based electronics can be difficult to monolithically integrate with other electronics technologies (Si-based CMOS digital, GaN-based high-power RF, *etc.*) because of the large lattice-mismatch between the materials families; there is continuing pressure to move to less toxic materials electronics (e.g. ROHS standard); finally, while III-V compound semiconductors have demonstrated $f_T$ and $f_{Max}$ cutoff frequencies in the low THz regime, even higher cutoff frequencies are needed for future commercial and DoD systems applications and it is not clear that III-V semiconductors can meet those demands [4]. New 2D materials with band gaps are predicted to have clear advantages for these applications, if the parasitic resistances can be reduced sufficiently. For all of these reasons, the time is ripe for investigation of new based 2D electronics materials such as graphyne that do not suffer from these limitations.

It is known that graphyne has allotropes which are planar, and the predicted electronic properties hold interest for high speed transport of carriers [5] – [9]. Growth and preparation of the material for further exploitation in electronic devices, for example, require various preparation methods.

Graphyne comes in allotropes which are planar and display bandgaps appropriate for semiconductor electronics. The various allotropes of graphyne are hexagonal honeycomb lattices, but with their benzene rings connected by acetylene triple C-C carbon-carbon bonds. This makes for extra complicated requirements for not only performing density functional calculations (DFT), but also in their physical preparation in a laboratory environment. Bandstructure will be shown in this letter to not be gapless as in graphene with its characteristic Dirac type bandstructure with its relativistic analogy to low mass particles, but parabolic type with its more massive characteristic of semiconductor action (and a bandgap).

There are several advantages to developing a new semiconductor material for electronics based on carbon related to graphene, because of the extensive prior work done on graphene growth and preparation and its electronics. The unique characteristic for graphyne is it consisting of single and triple acetylene bonds, allowing creation of a material with attractive thermal and electrical transport properties [5] – [7].

Graphyne exists in planar format, is predicted to have high mobility carriers, and impressive thermal/phonon properties. In addition, the sp-sp$^2$ multi-hybridized bonding character results in a number of possible graphyne allotropes, as shown in Figure 1. Wet chemistry methods have recently been developed to build carbon nanomaterials, including sp-sp$^2$ multi-hybridized graphyne systems, from the bottom up with atomic precision [8], [9]. Because of the large number of allotropes, graphyne provides a target-rich environment for synthetic organic efforts to have a positive impact on materials science. Our effort would be to find suitable ways to grow graphyne, create test structures, and determining electronic properties [10], proving or disproving the promise of this material, which has had almost no prior work [11] done on it to realize useful electronic devices, with the exception of a mechanochemistry method for γ-graphyne [12].

In order to attain the sp-sp$^2$ hybridization that is needed to grow graphyne, it was theoretically modeled [11] that the growth conditions must be in the carbon-deprived regime instead of the c-rich regime which is typical for graphene growth. Therefore, the growth conditions must be modified from the standard growth of graphene. Growth techniques variables that can be used to create sp-sp$^2$ bonding include growth temperature, addition of precursors, and post growth processing. Graphyne could be grown in using wet chemistry by coupling suitable small-molecule precursors via standard palladium-catalyzed cross-coupling reactions as well as via alkene and alkyne metathesis methods. Current homocoupling methods have already succeeded in bottom-up organic synthesis of graphene nanoribbons with precise edge structures, nitrogen and boron dopant placement, and the sp-sp$^2$ graphdiyne allotrope [6, 7]. Cross-coupling will extend the synthesis of these materials by allowing the use of multiple precursor molecules to fabricate one structure. Additionally, alkene and alkyne metathesis, [13] familiar to the synthetic organic community, have been severely underutilized in the synthesis of 2D carbon crystals such as graphyne.

In this letter we not only reexamine ordinary graphyne nanostructures which come in one-atom-thick planar sheets of sp and sp$^2$ bonded carbon atoms arranged in crystal lattice with a lattice of benzene rings connected by acetylene bonds, but also variants with more than single acetylene bonds. Graphyne consists of a mixed hybridization sp$^n$, where $1 < n < 2$; graphene and graphite (considered pure sp$^2$) and diamond (pure sp$^3$). Here we extend the definition of graphyne to the notation graphyne-n, where n counts the number of acetylene bonds. Ordinary graphyne is in this system, graphyne-1. The synthesis of graphdiyne , which is graphyne-2, was reported as a 1 mm film on a copper surface.

Furthermore, there are other applications and physical possibilities of control, as it appears likely in either pristine $\alpha$-, $\beta$-, or $\gamma$-graphyne or its substitutional doped versions with one or two N or B atoms per unit cell [14], removing or adding electrons compared to C atoms, that one can tune from a bandgap point in **k**-space to a Dirac point, using an electric field applied through a gate voltage (which moves the Fermi level). This is a feature available in no other 2D material, and would provide new electronic possibilities. Also optical [15] and magnetic [16] control is available, heretofore, of significantly larger magnitude than in graphene. For example, the introduction of a single atom vacancy per unit cell in graphyne, can induce a magnetic moment of $1.1 - 1.3$ $\mu_B$ in $\alpha$- and $\beta$-graphyne, and 1.8 $\mu_B$ in $\gamma$-graphyne.

One final point, and it is that although nanowires in the sense of nanotubes have been studied theoretically and experimentally for electronic device possibilities [17], [18], that is avoided that here because tube alignment is known to be a difficult challenge with substantial fabrication costs. Similarly, two terminal devices [19] – [21] are avoided to focus instead on transistor capability, where the largest impact on electronics will occur [22], [23]. This is why the search for new reproducible bandgap materials is so important.

## The Atomic Structures

Graphyne allotropes may be denoted by graphyne-n, where *n* indicates the number of carbon-carbon triple bonds in a link between two adjacent hexagonal benzene rings. Clearly, graphyne allotropes insert these triple bonds into graphene to create the new 2D material. Figure 1 shows atomic structures of the graphyne-1, graphyne-2, and graphyne-3, where n has been chosen to be $n = 1$, $n = 2$, and $n = 3$. Figure 2 gives graphyne-4 (left) and graphyne-5 (right).

### Structural Relaxation Electron Density Plot

Variable-cell structural optimization using Broyden–Fletcher–Goldfarb–Shanno (BFGS) quasi-newton algorithm was used to relax both the cell and the internal coordinates. The relaxed structures from both local density approximation (LDA) and generalized gradient approximation (GGA) calculations belong to P6The various/MMM space group and are shown in Fig 3. Volume of the LDA relaxed structure is 3.3% smaller than GGA.

Electron localization function (ELF) is a measure of the likelihood of finding an electron in the neighborhood space of a reference electron located at a given point and with the same spin. This has been plotted in Figure 4 for a GGA optimized structure, with C-C bond length in the benzene ring being 1.444 Å; C-C partial double-bond length being 1.415 Å, and C-C triple bond length being 1.219 Å.

### Bandstructure Calculations

One key technical issue for the development of these crystalline 2D semiconductors is to be able to predict accurate band structure and especially the band gap. The accuracy of the density functional theory (DFT) calculations strongly depends on the exchange-correlational functional, while the widely used local density approximation (LDA) and generalized gradient approximation (GGA) generally underestimate the band gap significantly. This is seen in Fig. 5. The various DFT exchange-correlational functionals have these known characteristics: LDA functional depends only on the local density; GGA functional depends on local density and its gradient; Meta-GGA functional depends on density, its gradient, and its second derivative; Hybrid functional mixes in Hartree-Fock exchange. In Schilfgaarde et al. [24], the LDA behavior is shown, as well as the random phase approximation (RPA) or GW approximation of Hedin [25]. Here G stands for Green's function, W is the screened coulomb interaction. Improvements came with Hybertsen and Louie [26] employing the LDA eigenfunctions to generate the GW self-energy $\Sigma = iGW$. The approach in [24], which is a modification of a full self-consistent $GW$ method (full SC$GW$), which they refer to as the quasiparticle self-consistent $GW$ (QS$GW$) method, results in accurate predictions of excited-state properties for a large number of weakly and moderately correlated materials. We won't belabor these points further, as long as one is aware of these intricacies of calculations.

We use quantum espresso, an integrated suite of open-source computer codes for electronic-structure calculations and materials modeling at the nanoscale. Quantum espresso is based on density-functional theory, plane waves basis set, and pseudopotentials, with many features to examine the ground-state calculations, including structural optimization, molecular dynamics, potential energy surfaces, electrochemistry and special boundary conditions, response properties, spectroscopic properties, quantum transport, and platforms. All the current calculations are performed using HPC computer Mustang at AFRL DSRC.

Calculated band structure of graphyne-1 for the γ allotrope, that is, γ-graphyne, using both LDA and GGA exchange-correlational functional is shown in Fig. 6, providing the energy ε(**k**) vs. **k** bandstructure through symmetry points in the Brillouin Zone, M, Γ and K. LDA and GGA calculated band structures show reasonable agreements, especially in close to the Fermi level. A direct band gap occurs at the M point in the Brillouin zone, with the gap of 0.49 eV for GGA and 0.411 eV for LDA.

For graphyne-2, again in the γ allotrope form, a variable-cell structural optimization using Broyden–Fletcher–Goldfarb–Shanno (BFGS) quasi-newton algorithm is used to relax both the cell and the internal coordinates. See Fig. 7. Both relaxed structures belong to P6/MMM space group. Volume of the LDA relaxed structure is 2.9% smaller than GGA. Calculated band structure of graphyne-2 using both LDA and GGA exchange-correlational functional is shown in Fig. 8. LDA and GGA calculated band structures show reasonable agreements, especially in close to the Fermi level. A direct band gap at the Γ point in the Brillouin zone, with the gap of 0.515 eV for GGA and 0.437 eV for LDA.

To put everything in perspective from and absolute value of the bandgap, examine the Table 1 reproduced from [27]. 40 different solid-state materials including C, Si, Ge, SiC, etc. are shown. LDA and GGA underestimate the band gap significantly; meta-GGA functional TPSS shows slight improvements; hybrid functional HSE improves the predicted band gap significantly. In this table the first two listings ME and MAE stand for, respectively, mean error (ME) and mean absolute error (MAE).

## Conclusions

In this treatment, two acetylene γ-allotropes of graphyne were studied and their bandstructures obtained. These two allotropes were the ones γ-graphyne-1 and γ-graphyne-2, with respectively, one and two acetylene triple bonded carbon atom pairs connecting the benzene ring lattice. Two standard DFT techniques were used to assess whether or not these two allotropes present finite bandgaps which may be usable for electronics. What was found were nominal bandgaps of about 0.5 eV, roughly half of what silicon presents. Because the LDA and GGA techniques used for the high performance supercomputer calculations, are known to most likely underestimate bandgaps as discussed, we are fairly certain that useful bandgaps for electronic devices such as transistors will be found when experimentally preparing these materials. Thus we expect the bandgaps $E_g$ to be $0.4 \lesssim E_g \lesssim 1.5\,eV$ .

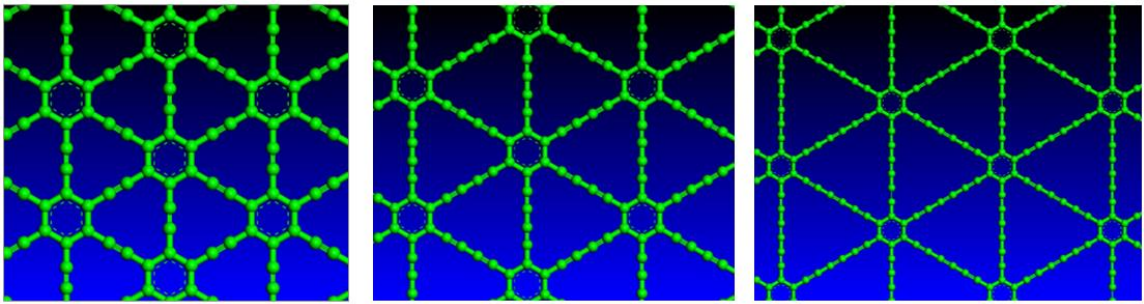

Figure 1: Top views of Graphyne-n varieties, which n indicates the number of carbon-carbon triple bonds in a link between two adjacent hexagons. Graphyne is graphyne-1 (left), graphdiyne is graphyne-2 (middle), and graphyne-3 (right).

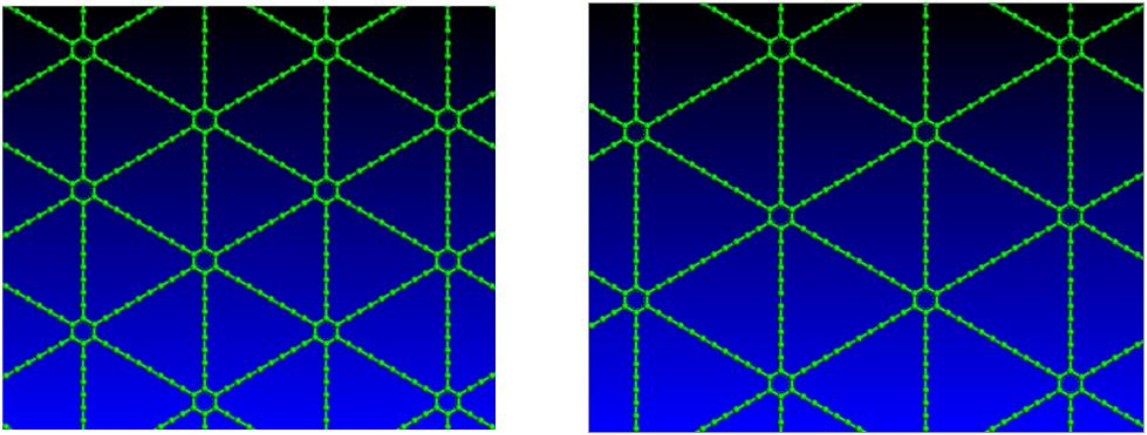

Figure 2: Graphyne-4 (left) and graphyne-5 (right) nanostructures.
.

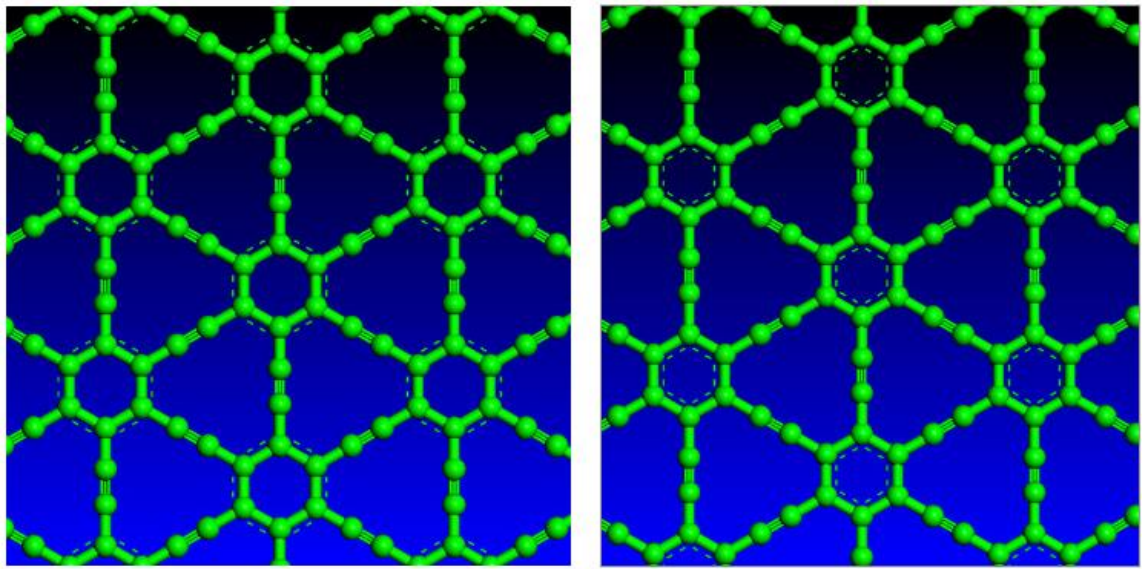

Figure 3. Optimized graphyne structures (graphyne-1) using LDA (left) and GGA (right).

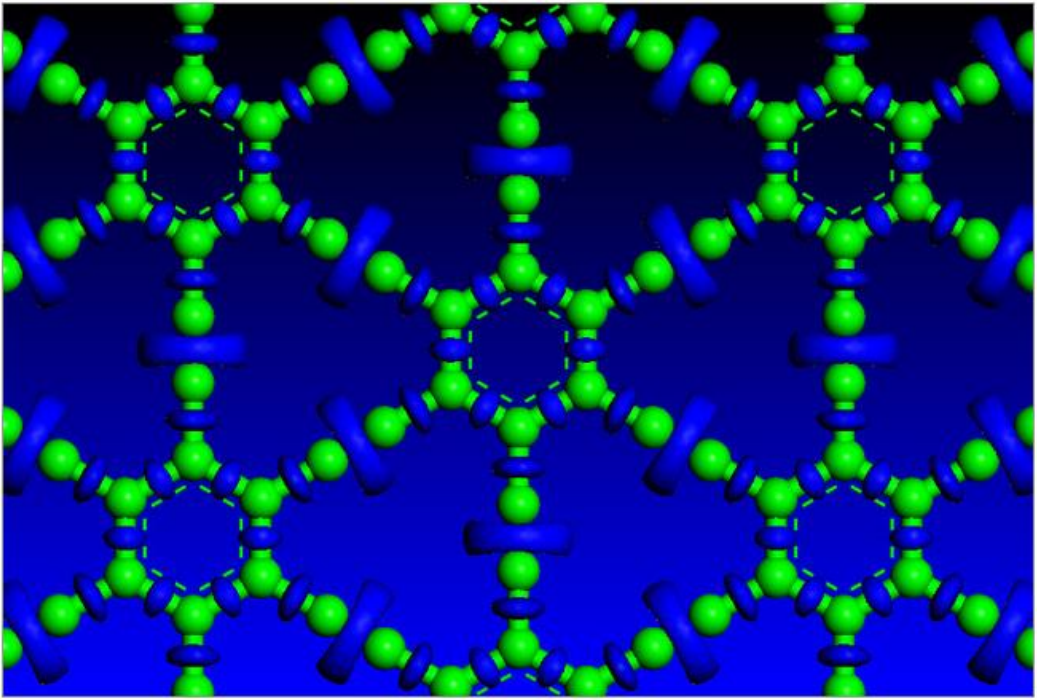

Figure 4. Electron localization function for a GGA optimized structure.

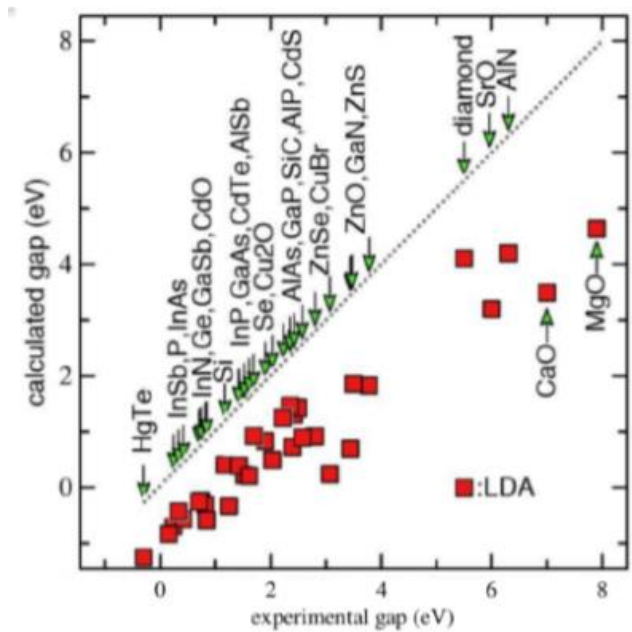

Figure 5. Infamous band gap problem for LDA and GGA.

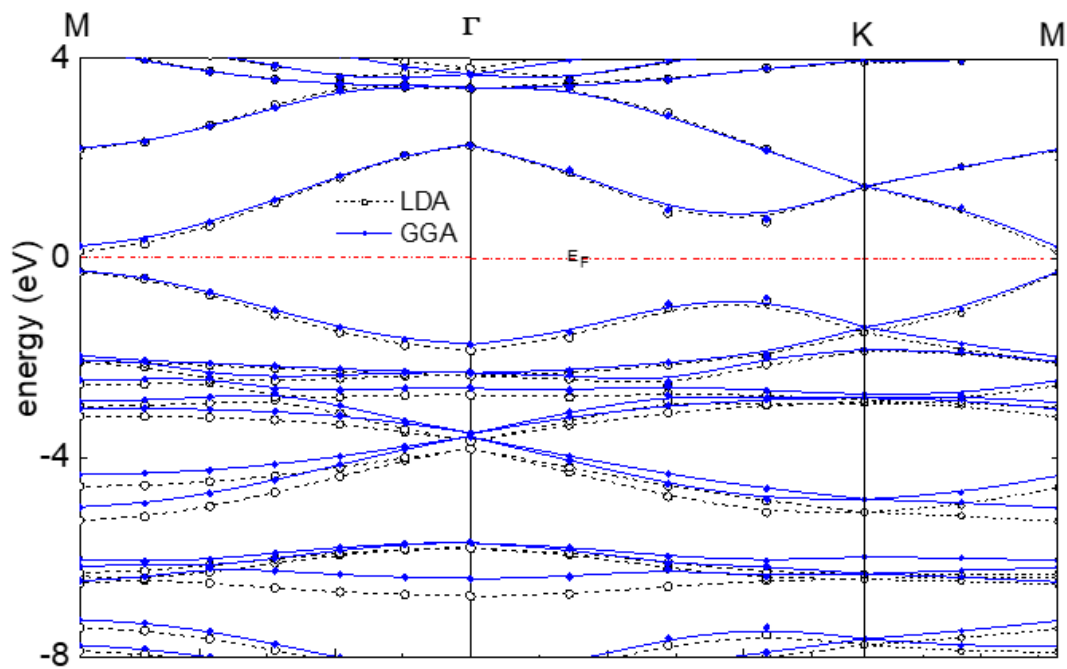

Figure 6. Calculated band structure of graphyne-1 using both LDA and GGA exchange-correlational functional.

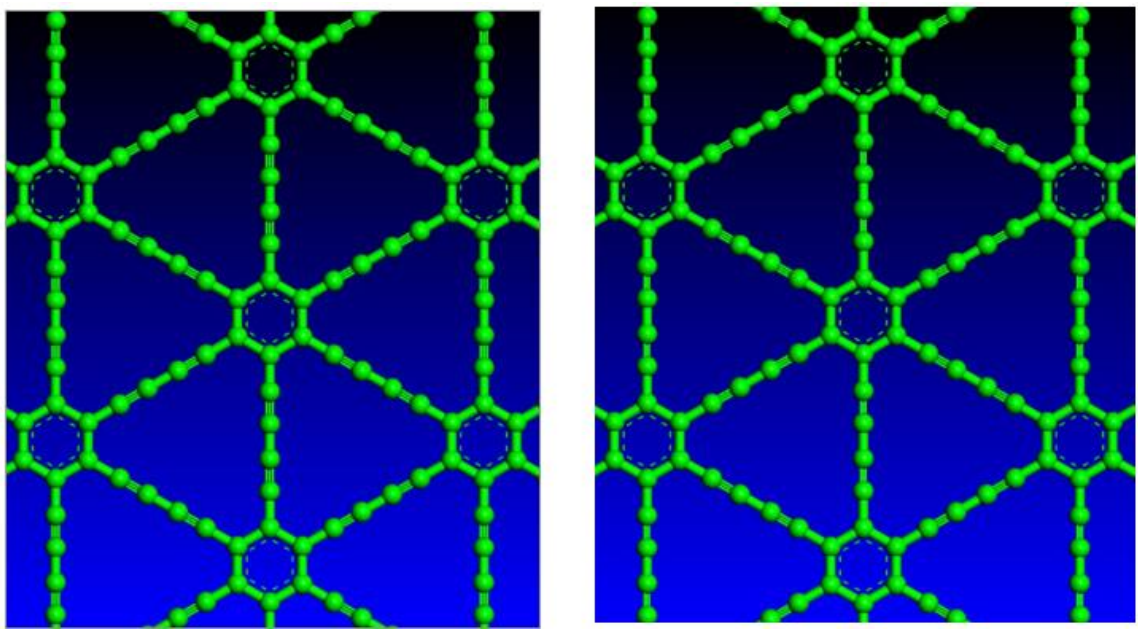

Figure 7. Optimized graphyne structures using LDA (left) and GGA (right). Variable-cell structural optimization using Broyden–Fletcher–Goldfarb–Shanno (BFGS) quasi-newton algorithm to relax both the cell and the internal coordinates.

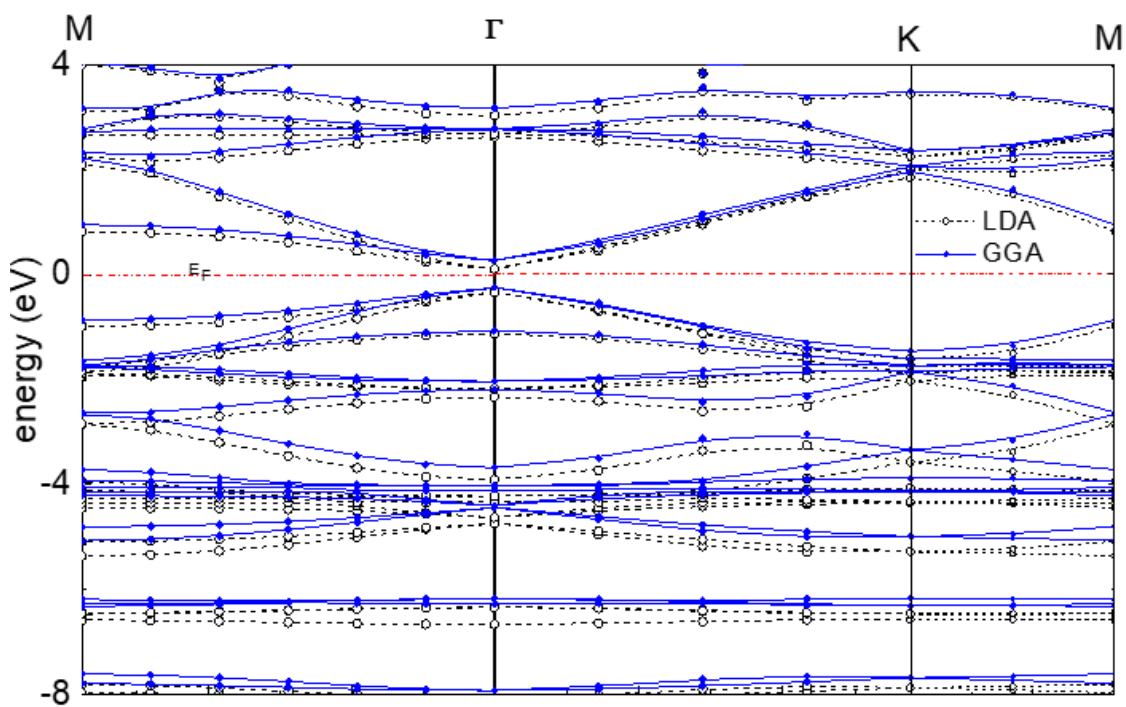

Figure 8. Calculated band structure of graphyne-2 using both LDA and GGA exchange-correlational functional.

Table 1: Bandgap error (eV) for 40 solid state materials

| Solid | LSDA | PBE | TPSS | HSE |
|---|---|---|---|---|
| ME[a] | −1.14 | −1.13 | −0.98 | −0.17 |
| MAE[b] | 1.14 | 1.13 | 0.98 | 0.26 |
| rms[c] | 1.24 | 1.25 | 1.12 | 0.34 |
| Max (+)[d] | ⋯ | ⋯ | ⋯ | 0.32 |
| Max (−)[e] | −2.30 | −2.88 | −2.66 | −0.72 |